\documentclass[preprint,12pt,a4paper]{article}
\usepackage[english]{babel}
\usepackage{graphics}

\begin{document}
\date{}
~\hfill {PAPER REFERENCE: A-III.2}
~\\ 
\begin{center}
{\large \bf  ELECTRONIC PROPERTIES OF\\ Mn-COMPOUNDS UNDER STRAIN}\\~\\
A. Debernardi,$^{1}$ M.~Peressi,
$^{\ast,1,2}$ and 
A.~Baldereschi$^{1,2,3}$
~\\ ~
(1) Istituto Nazionale di Fisica della Materia (INFM),\\
Research Unit of Trieste, via Beirut 4/2, I-34014 Trieste, Italy\\
(2) Department of Theoretical Physics, University of Trieste,\\
Strada Costiera 11, I-34014 Trieste, Italy\\
(3) Ecole Politechnique F\`ed\`eral de Lausanne,\\
PHB-Ecublens, CH-1015 Lausanne, Switzerland
\end{center}
~\\~\\
\hrule 
\begin{abstract}
We study the physical properties of MnAs under strain by using accurate 
first-principles pseudopotential calculations. 
Our results  provide new insight on the physics of strained multilayer that 
are grown  epitaxially on different lattice mismatched substrates 
 and which are presently of interest for spintronic applications. 
We compute the strain dependence of the structural parameters, electronic 
bands, density of states and magnetization. 
In the region of strain/stress that is easily directly accessible to 
measurements, the effects on these physical quantities are linear. We also 
address the case of  uniaxial stress inducing sizeable and 
strongly non linear effects on electronic and magnetic properties. 
\\~\\
{\it Keywors:} MnAs, strain, pseudopotential, magnetic properties,
electronic properties, epitaxy
%
\end{abstract}
\hrule
\footnotetext{$\ast$ Corresponding Author: 
Tel. +39 040 2240242; Fax. +39 040 224601; E-mail: peressi@ts.infn.it}
~\\~\\
{\bf 1. Introduction}\\~\\
The recent discovery of giant magneto-resistance
 has driven increasing interest
to the study of magnetic materials. The main target is the design
and the realization of new devices with materials whose
magnetic properties could be tailored in accordance to the needs
 of electronic industry.
Within the solid state community is rapidly becoming popular the
 word "spintronic" to denote electronic-like heterostructures
 where the relevant physical quantity
is the spin of the carriers
and its interactions with external magnetic fields
rather than the   charge of holes and electrons and
the associated electronic properties.
In this context the most promising materials are Mn compounds.

Although they  naturally occur in the hexagonal structure whose
symmetry is completely different with respect to the zincblende structure of
conventional semiconductors, high-quality epitaxial Mn compounds
can be grown on many semiconductor substrates, different for
composition and orientation.
Realistic cases include the growth 
on (111)-As-terminated GaAs or (111)Si~\cite{XXX-GaAs,XXX-Si}
where the MnAs $c$-axis is parallel to the (111) zincblende  substrate axis,
and on (001) GaAs or Si, with two possible different
orientations (usually referred as type-A and type-B)
of the MnAs $c$-axis with respect to the zig-zag
anion-cation bond chain underlying the substrate surface.
A recent experiment~\cite{XXX-ZnSe} has  proved
the possibility of growing ordered films (heterostructures)
of MnAs on ZnSe. The presence of a ZnSe buffer layer exclusively
stabilizes a particular phase (type-B) of MnAs;
at variance, the coexistence
of different phases in the magnetic epitaxial layers is experimentally
observed on GaAs substrates~\cite{XXX-GaAs}.

Due to the lattice mismatch with the
substrate, which is different according to its composition and orientation,
we expect the Mn-compound epilayers 
to grow in different kind of strained structures,
some of them with considerable deformation with respect 
to the original free-standing structure, and in general of reduced 
symmetry.

After a description of the structural, 
electronic and magnetic properties of MnAs in the free-standing 
hexagonal and zincblende structures, in this paper
 we address in particular  the effects on these
properties induced by a uniaxial stress along the $c$-axis
of the hexagonal structure.

~\\~\\{\bf 2. Computational details}\\~\\
Our calculations are performed within the
 spin-polarized density functional theory using ab-initio pseudopotentials
and plane-wave formulation.
For the exchange and correlation energy we use the generalized gradient 
approximation (GGA) with the Perdew-Wang 
parameterization~\cite{PW91}, that has been shown to
provide a good description of the metallic bond. 
From the comparative tests that we have performed using the 
local density approximation (LDA)
with the Perdew-Zunger parametrization
of the  exchange and correlation functional~\cite{PZ}, we found  that
GGA slightly improves the agreement of our results with respect to
the available experimental data (see Tab. 1). We have generated separable 
pseudopotentials in the Kleinman-Bylander form~\cite{KB}: the As  
pseudopotential has been generated according to the Martins-Troullier 
technique~\cite{TM}, while the Mn pseudopotential according 
to the Vanderbilt method~\cite{Van} 
which allows to have reliable results with a reasonable 
size of the plane-wave basis set (corresponding in the present case
 to a kinetic energy cutoff of 30 Ry). Reciprocal space 
integration in the  Brillouin zone 
is  performed using special k-point techniques~\cite{Chadi,MP}.
 A particular care in the choice of the sampling 
k-point set must be reserved to the system studied:
 at zero stress (for the free-standing hexagonal structure)
we  use a  ($12\times 12\times 8$) Monkhorst-Pack~\cite{MP}  
grid that corresponds to 336 k-points in the irreducible 
wedge of the first Brillouin zone, and  ensures an accurate
convergence of all the electronic properties studied. 
For the strained structures this set is modified 
accordingly.
The integration over the points of the Brillouin zone are performed 
with smearing techniques~\cite{smear} using a Gaussian broadening 
of 0.02 Ry. 

Deformations can be either introduced by constraining the structures in
some directions and allowing them to relax in the others (i.e., fixing
some components of the strain tensor  $\epsilon_{ij}$,
as it is the case
for epitaxial structures that must accommodate the lattice
mismatch with the substrate), or by applying a stress $\sigma_{ij}$
(external force).
We focus  here on the last case, {\em simulating the application of a uniaxial 
stress along the $c$ axis of the hexagonal structure}, here referred to as 
the $z$ direction. The correspondent equilibrium lattice structures 
must have \emph{vanishing stress} along the other 
directions, i.e. $\sigma_{zz}\ne 0$ and $\sigma_{xx}=\sigma_{yy}=0$. 
In practice in the numerical experiments  we consider several 
structures with different strain tensors $\epsilon_{ij}$, with 
$\epsilon_{xx}=\epsilon_{yy}\ne \epsilon_{zz}$,
and we compute the stress tensor $\sigma_{ij}$
according to the formalism introduced by 
Nielsen and Martin.~\cite{Stress_the,Stress_the1} 
 The structures
can be considered optimized  if the residual
stress in the $xy$ plane is within $\approx$1kBar. 
We point out that in general this condition is
{\em not} equivalent to the one of volume conservation, i.e. zero-trace
strain tensor, and requires a larger computational effort.

~\\~\\{\bf 3. Results}\\~\\
{\em 3.1 Free-standing unstrained structures}\\
In normal free-standing conditions MnAs is a ferromagnetic metal
with the NiAs structure, i.e. 
hexagonal lattice with four atoms in the unit cell, which is described
by the parameters $a$ (corner of the hexagon) and $c$ (Fig. 1). 
The zincblende structure is
experimentally observed only for the related alloy
Ga$_{1-x}$Mn$_x$As with the extreme diluted concentration
of Mn atoms, $x<0.07$. We will address here also the study 
of this structure, for a comparison with the hexagonal one.
In Tab. 1 we summarize the results for the structural
and magnetic properties obtained with our pseudopotentials
with GGA, together with  those obtained
within LDA, and also by other authors using full-potential
linearized augmented plane-wave techniques (FLAPW)~\cite{Picon}.
The comparison with the available experimental data completes the Table.

Comparing zincblende and hexagonal MnAs, we found that
the last one is more dense and  more stable (by $\sim$0.7 eV);
the increase in the atomic distance in the zincblende structure
is accompanied by an increase in the total magnetization $M_{tot}$, which
goes from 2.80 $\mu_B$ to 3.84 $\mu_B$, to 
be compared with the limiting value of 4 $\mu_B$ per MnAs pair
for atoms at infinite distance.
We compute both the total and the absolute magnetization,
$M_{tot}={1\over\Omega_{cell}}
\int \left(\rho^\uparrow({\bf r})-\rho^\downarrow({\bf r})\right)d{\bf r}$ and
$M_{abs}={1\over\Omega_{cell}}
\int \left|\rho^\uparrow({\bf r})-\rho^\downarrow({\bf r})\right|d{\bf r}$.

The two structures are very different as far as the electronic properties
is concerned: at variance with the hexagonal phase, the zincblende
MnAs is half-metallic, as  we show in Fig. 2. 
From  the density of states (DOS) for the majority and minority spin 
projected onto the different atomic states (not shown here)
 it is evident the atomic origin ($d$ states of Mn) of the flat bands
in the two spin orientations.
Our results are  in agreement with some obtained from previous pseudopotential 
calculations.~\cite{Savito-zb}

~\\~\\{\em 3.2 Strained structures for applied uniaxial stress}\\
The uniaxial stress along the $c$ axis 
 does not cause any reduction of the symmetry of the structure,
which remains hexagonal but with a $c/a$ ratio different from the
free-standing one. 
For uniaxial stresses up to about 5.2 GPa the effects on the
structural parameters, electronic structure and magnetic properties are
almost linear. For the electronic structure we report the bands
for the unstrained and for one strained configuration (Fig. 3, solid 
and dashed lines), and the total DOS projected 
onto the different atomic states for the same configurations and for a third 
one with intermediate strain
configuration (Fig. 4). No dramatic changes are observed, 
only a progressively filling of the DOS around the Fermi energy.
The total magnetization reduces linearly with the stress
from 2.80 $\mu_B$ to 2.65 $\mu_B$.

Sizeable changes occur instead for higher  uniaxial stress.
At about 7.6  GPa the effect of the applied stress
on the structural properties is no longer linear:
it can be seen in terms of induced strain, or volume change
 (which amounts to $-$10\%), or
reduction of $c/a$ ratio (of about $-20$\%, thus corresponding to 
a strong cell deformation). 
Also the effects on the electronic and magnetic properties are important.
The band structure  (Fig. 5) looks quite different from the 
free-standing and from the low-stress region: the bands
originating from the $d$-states of Mn approach the Fermi
level, going towards higher energy for the majority spin,
and towards lower energy for the minority spin. The net effect is
a dramatic reduction of the magnetization, which goes to 0.26 $\mu_B$.
Although further investigation is needed to better understand the  microscopic
origin of such effect, our preliminary results seem to indicate that 
the magnetization reduction is related not simply to a sizeable 
compression, but, more important, to a strong deformation of the structure.

~\\~\\{\bf 4. Summary}\\~\\
We calculate the variation of the structural parameters, electronic 
bands, density of states and magnetization of MnAs under the application of
a uniaxial stress along the $c$ axis of the hexagonal structure. 
We found a region of strain/stress where the effects 
are linear and rather small, but also a region of sizeable and 
strongly non linear effects on electronic and magnetic properties.

~\\~\\{\bf 5. Acknowledgments}\\~\\
Calculations in this work have been done using the PWSCF
 package~\cite{PWSCF}.
We acknowledge the Istituto Nazionale di Fisica della Materia for
 the ``Iniziativa Trasversale di Calcolo Parallelo''.
We  acknowledge D.~Vanderbilt to have kindly provided us the latest 
version of his pseudopotential code.

\newpage
\begin{table}
\begin{center}
Table 1\\~\\
\begin{tabular}{|l|c|c|c|c|c|c|} 
\hline
&crystal str.& E$_{xc}$ & a$_{L}$ & $c/a$ & $M_{tot}$ & $M_{abs}$  \\ 
&            &        &  (\AA)   &       & ($\mu_B$)  & ($\mu_B$) \\
\hline
Pseudo    & HEX  & LSDA & 3.51   & 1.49   & 2.15   & 2.39 \\
Pseudo    & HEX  & GGA  & 3.64   & 1.50   & 2.80   & 3.13 \\  
FLAPW$^a$ & HEX  & LSDA & 3.487  & 1.49   & 1.91   &  -  \\
FLAPW$^a$ & HEX  & GGA  & 3.704  & 1.49   & 3.15   &  -  \\
Exp.      & HEX  &      & 3.7    &1.54    & 3.4  &  - \\
\hline
FLAPW$^a$ & FCC  & GGA  & 5.643  &    -   & 3.83   &  -  \\
Pseudo    & FCC  & GGA  & 5.69   &    -   & 3.84   & 4.24 \\
\hline
\end{tabular} \\ 
$^a$ from Ref.\protect~\cite{Picon} 
\end{center}
\caption{
Structural parameters and magnetization for MnAs in the FCC and HEX phases
compared with other calculations and with available experiments.
The pseudopotential results, both with LDA and GGA, are those obtained
in the present work.
}
\end{table}
~\vfill
\newpage
\begin{figure}
\caption{
The NiAs structure.}\end{figure}

\begin{figure}
\caption{
Majority and minority bands for the half-metallic zincblende MnAs.
The energy zero is  set to the Fermi energy. 
}\end{figure}

\begin{figure}
\caption{
Majority and minority bands for the ferromagnetic MnAs in the
unstrained structure and in the structure corresponding to 
 an applied uniaxial stress on $c$ of 5.2 GPa.
The energy zero is  set to the Fermi energy. 
}\end{figure}

\begin{figure}
\caption{
Total density of states  projected 
onto the different atomic states for the unstrained hexagonal MnAs,
and for two different strain conditions corresponding to an applied
uniaxial stress on $c$ respectively of  2.6 GPa and 5.2 GPa.
}\end{figure}

\begin{figure}
\caption{
Majority and minority bands for the ferromagnetic MnAs in the
unstrained structure and in the structure corresponding to 
 an applied uniaxial stress on $c$ of 7.6 GPa.
The zero is arbitrary set to the Fermi energy. 
}\end {figure}
\end{document}